\documentclass[a4paper,12pt]{article}

\usepackage{amsthm}
\usepackage{amsbsy}
\usepackage{epic}
\usepackage{graphicx}
\usepackage{graphics}
\usepackage{amssymb}
\usepackage{psfrag}
\usepackage{amsmath}
\usepackage{epsfig}
\usepackage{array}
\usepackage[round]{natbib}
\usepackage{algorithm}
\begin{document}
\begin{center}
{\bf{Efficient Bayesian Analysis of Multiple Changepoint Models with 
Dependence across Segments}}\\
Paul Fearnhead and Zhen Liu \\
Department of Mathematics and Statistics, Lancaster University
\end{center}
{\bf{Summary:}} We consider Bayesian analysis of a class of multiple changepoint models. While there are a variety of efficient ways to analyse these models if the parameters associated with each segment are independent, there are few general approaches for models where the parameters are dependent. Under the assumption that the dependence is Markov, we propose an efficient online algorithm for sampling from an approximation to the posterior distribution of the number and position of the changepoints. In a simulation study, we show that the approximation introduced is negligible. We illustrate the power of our approach through fitting piecewise polynomial models to data, under a model which allows for either continuity or discontinuity of the underlying curve at each changepoint. This method is competitive with, or out-performs, other methods for inferring curves from noisy data; and uniquely it allows for inference of the locations of discontinuities in the underlying curve.  \\
\section{Introduction}

Changepoint models are commonly used for time-series, to allow for abrupt changes in the underlying model or structure for the data. Some example applications areas include  genetics \cite[]{Liu/Lawrence:1999,McVean/Myers:2004}, environmental time-series \cite[]{Dobigeon:2007,Seidou:2007}, and signal processing \cite[]{Punskaya:2002}, amongst many others. 

We consider Bayesian inference for changepoint models. Existing methods for such inference are either based on MCMC approaches \cite[e.g.][]{Stephens:1994,Chib:1996,Chib:1998,Lavielle/Lebarbier:2001}, or methods for direct simulation from the posterior \cite[see e.g.][]{Yao:1984,Barry/Hartigan:1992,Liu/Lawrence:1999,Fearnhead:2008}. The methods for direct simulation have the advantage over MCMC of producing iid draws from the posterior, and they can also be implemented efficiently so that their computational cost is linear in the number of observations \cite[]{Fearnhead/Liu:2007}. However they are limited in terms of the class of models that can be considered. If we call the period of time between two successive changepoints a segment, then direct simulation methods require the parameters associated with each segment to be independent of each other, and that the marginal likelihood for the data within each segment can be calculated analytically \cite[or numerically, see][]{Fearnhead:2006SC}. 

One implementation of the direct simulation methods is based on solving filtering recursions \cite[]{Fearnhead/Liu:2007}. We process the observations one at a time, and when processing the observation at a time $t$ say, we calculate the posterior distribution of the time of the most recent changepoint prior to $t$. Here, we extend this direct simulation methods to models where there is dependence across segments. We assume that the dependence is Markov, so that parameters in the current segment depend only on the parameters in the previous segment. The assumption of dependence across segments greatly increases the complexity of calculating the posterior distribution, and to develop a computationally efficient algorithm we introduce a simple approximation. At time $t$ we approximate the distribution of the parameters associated with a new segment, conditional on a changepoint at $t$. While this 
conditional distribution is a mixture distribution, with the number of terms in the mixture increasing exponentially with $t$, we approximate the mixture by a single distribution. This approximation leads to an efficient algorithm, but one that produces iid samples from an approximation to the posterior distribution of interest. 

We demonstrate our new method on the problem of fitting piece-wise polynomial models. Here dependence across segments arises due to assumptions of continuity of the underlying curve. Existing methods for this problem include the MCMC methods of \cite{Denison/Mallick/Smith:1998} and \cite{DiMatteo/Genovese/Kass:2001}, who also sample from an approximation to the posterior of interest, from approximating the marginal likelihood associated with each segment. Our model extends existing models that are considered, by allowing for the possibility of either continuity or discontinuity of the curve at each changepoint. Our approach also allows for online analysis of time-series.

The outline of the paper is as follows. Firstly we introduce the class of changepoint models we consider. Then in Section 3 we develop out algorithm for Bayesian inference for these models. Section 4 then analyses the resulting algorithm for the specific application of 
fitting piece-wise polynomial models. We first show that the approximation introduces negligible error when analysing simulated data from the true model. We also compare the resulting method with both wavelet-based methods and the MCMC method of \cite{Denison/Mallick/Smith:1998}, and look at the power of the method for detecting discontinuities in the underlying signal. Section 5 applies our method to analysing well-log data. Here the focus of inference is in detecting changepoints where the  underlying signal is discontinuous. Finally the paper ends with a discussion.

\section{Changepoint model} \label{sec:model}
We consider the following hierarchical model for observations $y_{1:n}=(y_1,\ldots,y_n)$.  Firstly we introduce a model for the number, $l$, and position,
$0<\tau_1<\cdots<\tau_l<n$, of the changepoints.  This is based on a distribution for 
the distance between two successive changepoints
\begin{eqnarray}
\label{eqn:point_process}
p(\tau_k-\tau_{k-1}=d) = g(d),
\end{eqnarray}
for some discrete distribution $g(\cdot)$ on the positive integers. We define $\tau_0=0$ and 
$\tau_{l+1}=n$, and we let $G(s)=\sum_{d=1}^s g(d)$ be the corresponding cumulative distribution 
function. We assume independence of the distance between different pairs of successive 
changepoints, so that the joint probability of $l$ specific changepoints is
\[
 \Pr(\tau_1,\ldots,\tau_l)=\left (\prod_{k=1}^l g(\tau_k-\tau_{k-1})\right )(1-G(n-\tau_l)).
\]

The changepoints split the data into $l+1$ segments, with the $k$th segment containing 
observations $\mathbf{y}_{\tau_k+1:\tau_{k+1}}$, for $k=0, \ldots, l$.
For segment $k$ we associate a model $M_k$ and a vector of parameters $\theta_k$. The model is drawn from a  finite set of possible models, $\mathcal{M}$ and we assume that there is independence of the choice of model across different segments. \cite[Extension to the case where the model of a segment depends on the model of the previous segment is possible, see][]{Fearnhead/Vasileiou:2009}.

For $k\geq1$ we allow the distribution of $\theta_k$ to depend on the position of segment $k-1$, $\tau_{k-1}$ and $\tau_k$, and its parameter $\theta_{k-1}$.  Thus we  have 
that the conditional probability of the model and parameters for the segment can be factorised as
\[
\Pr(M_k=m)p_{m}(\theta_k|\theta_{k-1}\tau_k,\tau_{k-1}).
\]
For the first segment we assume a prior for $\theta_0$. Note that this framework includes models where there are common parameters across segments. In this case some components of $\theta_k$ are equal to the equivalent components of $\theta_{k-1}$ and the conditional probability $p_{m}(\theta_k|\theta_{k-1}\tau_k,\tau_{k-1})$ in only non-zero for parameter combinations that obey this constraint.

Given a segment defined by changepoints at positions $s$ and $t$, and with model $m$ and parameter $\theta$ we have a likelihood model
\begin{eqnarray}
\label{eqn:local_model}
p_{m}(\mathbf{y}_{s+1:t}|\theta).
\end{eqnarray}
We assume that conditional on the changepoints, segment models and parameters, the observations within each segment are independent of each other.

Finally we assume that there exists a family of conjugate priors for $\theta$, $p_m(\theta|\zeta)$. Thus for all $m$, $\zeta$ and $y_t$ and $s,t$, we can calculate
\begin{eqnarray}
\label{eqn:likelihood} 
P_s(t,m,\zeta)
&=& \int p_{m}({y}_{t}|\theta,s)p_{m}(\theta|\zeta)
\mathrm{d}\theta,
\end{eqnarray}
where 
\[
p_{m}({y}_{t}|\theta,s)=\frac{p_{m}(\mathbf{y}_{s+1:t}|\theta)}{p_{m}(\mathbf{y}_{s+1:t-1}|\theta)}
 \]
is the probability density of $y_t$ given a segment that started with observation $y_{s+1}$.
Furthermore, conjugacy imples that there exists a $\zeta'$ such that
\begin{equation} \label{eq:conj}
p_m(\theta|\zeta')\propto p_{m}({y}_{t}|\theta,s)p_{m}(\theta|\zeta),
\end{equation}
where the constant of proportionality is defined so that the right-hand side integrates to 1 (with respect to $\theta$).
We denote the value of $\zeta'$ defined by (\ref{eq:conj}) by an update function $u_s$:
\begin{equation} \label{eq:parup}
\zeta'=u_s(t,m,\zeta).
\end{equation}
This update function (and hence $\zeta'$) will depend on the data $y_{s+1:t}$.

We now give an example of such a changepoint model, which will be used throughout the paper to demonstrate and make concrete the ideas we present.

{\bf{Example: Piecewise Polynomial Regression}}

We consider filtering a piecewise polynomial regression model to bi-variate data $(x_i,y_i)$
for $i=1, \ldots, n$, with the data ordered so that $x_1<x_2<\cdots<x_n$. For concreteness
we will focus on piecewise quadratic models, but the extension to polynomials of different
order is straightforward.

If the observations $\mathbf{y}_{s+1:t}$ are in the $k$th segment, we 
specify the model of (\ref{eqn:local_model}) by:
\begin{eqnarray}
\label{eqn:poly}
\mathbf{y}_{s+1:t} = \mathbf{H}_k\beta_k+\varepsilon_k,
\end{eqnarray}
where the design matrix $\mathbf{H}_k$ is of form
\begin{eqnarray*}
\mathbf{H}_k =
\left (
\begin{array}{ccc}
1 & 0 & 0 \\
1 & x_{s+2}-x_{s+1} & (x_{s+2}-x_{s+1})^2  \\
\vdots & \vdots  & \vdots \\
1 & x_{t}-x_{s+1} & (x_{t}-x_{s+1})^2
\end{array}
\right ), 
\end{eqnarray*}
$\varepsilon_k$ is a vector of noises that are independently drawn from a $N(0, \sigma^2)$ 
distribution, and $\beta_k=(\beta_{k,0},\beta_{k,1},\beta_{k,2})$ is a vector-valued regression 
parameter. 

For simplicity, we model the distance between successive changepoints as geometric with mean $1/p$, so $g(d)=p(1-p)^{d-1}$. For each
segment except the first we allow for one of two models: $M=1$ refers to the underlying curve being discontinuous at the changepoint that starts the segment, and $M=2$ refers to the curve being continuous at this changepoint. Our prior is that the model of each segment is equally likely to be either possibility. Note that if $M=2$ then $\beta_{k,0}$ will be determined by the length and parameters of the previous segment. 

We assume that $\sigma^2$ is common to all the segments. However, to be consistent with the above 
framework, we introduce $\sigma_k^2$ to denote its value in the $k$th segment. 
Thus we have that $\theta_k=(\sigma_k,\beta_k)$ and $\theta_k$ depends on $\theta_{k-1}$ as $\sigma_k=\sigma_{k-1}$, and if $M=2$ through the dependence of $\beta_{k,0}$ on $\beta_{k-1}$.
 
We use the following standard conjugate priors for the variance $\sigma_k^2$ and the regression parameter
$\beta_k$ for both $M=1,2$:
\begin{eqnarray}
\sigma^2_k &\sim& \mbox{IG}(\nu/2,\gamma/2), \nonumber \\
\beta_{k}|\sigma^2_k &\sim& \mbox{N}(\mathbf{\mu},\sigma^2_k \mathbf{D}), \label{eq:p2}
\end{eqnarray}
where $\mbox{IG}$ denotes the inverse Gamma distribution and $\mbox{N}$ denotes the Gaussian 
distribution. With the notation above, we have $\zeta=(\nu,\gamma,\mu,\mathbf{D})$. For the first segment, for which $M_0=1$, we have prior parameter $\zeta_{0,1}=(\nu_0,\gamma_0,\mathbf{0},\mathbf{D}_0)$, with $\mathbf{D}_0=\mbox{diag}(\delta_0,\delta_1,\delta_2)$. For a future segment $k$ with $M_k=1$, we have the distribution for $\beta_k$ given by (\ref{eq:p2}) with $\mathbf{\mu}=(0,0,0)$ and $\mathbf{D}=\mathbf{D}_0$. For a segment $k$ with $M_k=2$, and the previous segment starting with observation $x_{r+1}$ and ends with observation $x_s$, the distribution for $\beta_k$ is given by (\ref{eq:p2}) with $\mathbf{\mu}=(\beta_{k-1,0}+\Delta\beta_{k-1,1}+\Delta^2\beta_{k-1,2},0,0)$, where $\Delta=(x_{s+1}-x_{r+1})$, and $\mathbf{D}=\mbox{diag}(0,\delta_1,\delta_2)$. This prior distribution ensures continuity of the underlying curve.

We can calculate $P_s(t,m,\zeta)$ and $u_s(t,m,\zeta)$ (see Equations \ref{eqn:likelihood} and \ref{eq:parup}) using standard
updates for dynamic linear models \cite[]{West/Harrison:1989}; details are given in the Appendix. Given the changepoint positions and segment models, 
we have a linear model for our data, and due to the choice of priors we can simulate directly from the posterior distribution of the parameters. The difficulty with Bayesian inference for this model is due to the intractability of the posterior distribution for changepoint positions and segment models.

\section{Approximate Inference}

We now describe our method for drawing, approximately, from the posterior distribution of the number and position of changements, and model and parameter values for each segment. The approach is based on recursive filtering and smoothing algorithms, which we will describe in turn. Throughout our description we will introduce a (potentially artificial) time, with observation $y_t$ arriving at time $t$. For ease of presentation it will be useful to refer to the model and parameter values associated with the segment to which $y_t$ belongs. Hence, for the rest of the paper we will slightly change notation, with $\theta_t$ and $M_t$ refering to the parameter and model value at this time $t$. That is we will {\em {subscript by time rather than by segment}}. We also introduce a new variable, $C_t$, which will denote the position of the most recent changepoint prior to time $t$.

\subsection{Filtering Algorithm}
\label{sec:filter}

To simplify the following exposition we will first derive the filtering algorithm for the case of a geometric segment length, $g(d)=p(1-p)^{d-1}$. Presentation of the algorithms we derive (Algorithms 1 and 2), include the details for a general segment length distribution.

First note that $(C_t,M_t,\theta_t)$ are a Markov process; and in particular the marginal dynamics for $C_t,M_t$ are given by
\begin{eqnarray*}
p(C_{t+1}=j,M_{t+1}=m|C_{t}=i,M_t=m')=\left \{ \begin{array}{ll}
1-p & \textrm{if $j=i$ and $m=m'$}, \\
p\Pr(M=m) & \textrm{if $j=t$, $m\in\mathcal{M}$}, \\
0 & \mbox{otherwise}.
\end{array} \right. 
\end{eqnarray*} 
The top probability refers to there not being a changepoint
between $y_t$ and $y_{t+1}$, and the middle probability refers to the event that there is.

Now we wish to recursively approximate
\[
p(C_t,M_t,\theta_t|y_{1:t})=p(C_t,M_t|y_{1:t})p(\theta_t|y_{1:t},C_t,M_t).
\] 
The first term on the right-hand side is a discrete
distribution, and  we approximate $p(C_t=s,M_t=m|y_{1:t})\approx w_t^{(s,m)}$.
Whereas for given $C_t=s$ and $M_t=m$ we will approximate $p(\theta_t|y_{1:t},C_t,M_t)$ by
$p_m(\theta_t|\zeta^{(s,m)}_t)$, for some $\zeta^{(s,m)}_t$. 
Our approximation is specified by the set of probabilities $w_t^{(s,m)}$ and parameters $\zeta_t^{(s,m)}$ for 
$s=0,\ldots,t-1$ and $m\in\mathcal{M}$ the set of possible models.

We initiate our algorithm using the model prior, with $w_0^{(0,m)}=\Pr(M=m)$ for $m\in\mathcal{M}$, and prior for the parameters $\zeta_0^{(0,m)}=\zeta_{0,m}$. For $t=1,\ldots,n$ we have the following set of recursions. Firstly for $s\in\{0,\ldots,t-1\}$, $C_{t+1}=s$ means that there is no changepoint at time $t$. Thus we have $M_{t+1}=M_{t}$ and $\theta_{t+1}=\theta_{t}$, and
\begin{eqnarray*}
\lefteqn{p(C_{t+1}=s,M_{t+1}=m,\theta_{t+1}=\theta|y_{1:t+1})=}\\
& & K p(C_{t}=s,M_{t}=m,\theta_{t}=\theta|y_{1:t})\Pr(C_{t+1}=s|C_{t}=s)p_m(y_{t+1}|\theta),
\end{eqnarray*}
for some normalising constant $K$.
Now we substite our approximation, $p(C_{t}=s,M_{t}=m,\theta_{t+1}|y_{1:t})\approx w_t^{(s,m)}p(\theta|\zeta_t^{(s,m)})$. Integrating with respect to $\theta$ gives $\Pr(C_t=s,M_t=m|y_{1:t+1})$, and thus
\[
w_{t+1}^{s,m}=K w_t^{s,m}(1-p)P_s\left(t+1,m,\zeta_{t}^{(s,m)}\right).
\]
While, using the updates for the conjugate distribution for $\theta$ we get 
\[
p(\theta_{t+1}|y_{1:t+1},C_{t+1}=s,M_{t+1}=m)\propto p_m(\theta_{t+1}|\zeta_{t}^{(s,m)})p_m(y_{t+1}|\theta_{t+1})
=p_m(\theta_{t+1}|\zeta_{t+1}^{(s,m)}),
\]
for $\zeta_{t+1}^{(s,m)}=u_s(t+1,m,\zeta_t^{(s,m)})$.

Now consider $C_{t+1}=t$. This corresponds to a changepoint at time $t$, and $C_{t}$ can take any value in $\{0,\ldots,t-1\}$. 
We derive an approximate recursion by considering
\begin{eqnarray*}
\lefteqn{p(C_{t+1}=t,M_{t+1}=m,\theta_{t+1}|y_{1:t})=}\\ 
& & \sum_{s=0}^{t-1}\sum_{m'\in\mathcal{M}} 
p(C_{t}=s,M_{t}=m',\theta_{t}|y_{1:t})\Pr(C_{t+1}=t|C_{t}=s)\Pr(M=m)p(\theta_{t+1}|\theta_t,t,s),
\end{eqnarray*}
where $p(\theta_{t+1}|\theta_t,t,s)$ denotes the conditional distribution of $\theta_{t+1}$ given $\theta_t$ and that the previous segment contained observations $y_{s+1:t}$. Now, substituing our approximations to $p(C_{t}=s,M_{t}=m',\theta_{t}|y_{1:t})$ we have
\begin{equation}\label{eq:pred}
p(\theta_{t+1}|y_{1:t},C_{t+1}=t,M_{t+1}=m)\propto \sum_{s=0}^{t-1}\sum_{m'\in\mathcal{M}} p w_t^{(s,m')}p_m(\theta_t|\zeta_t^{(s,m')})p(\theta_{t+1}|\theta_t,t,s).
\end{equation}
Our approach is to approximate this by $p_m(\theta_{t+1}|\zeta_{t}^{(t,m)})$ for some suitable choice of $\zeta_{t}^{(t,m)}$. Thus as
\begin{eqnarray*}
\lefteqn{p(C_{t+1}=t,M_{t+1}=m,\theta_{t+1}|y_{1:t+1})= K p(C_{t+1}=t,M_{t+1}=m,\theta_{t+1}|y_{1:t})p_m(y_{t+1}|\theta_{t+1})}\\
&=& Kp(C_{t+1}=t,M_{t+1}=m|y_{1:t})p(\theta_{t+1}|y_{1:t},C_{t+1}=t,M_{t+1}=m)p_m(y_{t+1}|\theta_{t+1}),
\end{eqnarray*}
we get the approximate recursion
\[
w_{t+1}^{(t,m)}=K\Pr(M=m)P_t(t+1,m,\zeta_{t}^{(t,m)}) \sum_{s=0}^{t-1}\sum_{m'\in\mathcal{M}} w_t^{(s,m')}p, 
\]
and $\zeta_{t+1}^{(t,m)}=u_t(t+1,m,\zeta_t^{(t,m)})$. 

Note that the only approximation in our filtering recursions is in the approximation of (\ref{eq:pred}). There are various ways of choosing $\zeta_t^{(t,m)}$ for this approximation, and in practice we use a simple method of moments approach (see below). Note that this approximation is required to avoid the exponentially increasing computational cost of the exact filtering recursions. Similar approximations have been used in the Generalised Pseudo-Bayes algorithm \cite[]{Tugnait:1982}, or the Interacting Multiple Model filter \cite[]{Blom:1988}. 

The full filtering algorithm, allowing for a general distribution of segment lengths and prior distribution for models is described in Algorithm \ref{alg:filter}.
\begin{algorithm}
    \caption{Filtering Algorithm}
    \label{alg:filter}
\begin{description}
\item[Initiate] Set ${w}_1^{(0,m)}=\Pr(M=m)P_s(1,m,\zeta_0^{(0,m)})$ and $\zeta_1^{(0,m)}=u_s(1,m,\zeta_0^{(0,m)})$ for $m\in\mathcal{M}$. Normalise weights, $w_1^{(0,m)}$ and let $t=1$.
\item[While $t<n$] 
\begin{itemize}
\item[(i)] For $s=0,\ldots,t-1$ and $m\in\mathcal{M}$, set 
\[
{w}_{t+1}^{(s,m)}=\frac{1-G(t+1-s)}{1-G(t-s)} w_t^{(s,m)}P_s\left(t+1,m,\zeta_t^{(s,m)}\right),
\]
 and $\zeta_{t+1}^{(s,m)}=u_s(t+1,m,\zeta_{t}^{(s,m)}$.
 \item[(ii)] For $m\in\mathcal{M}$, calculate $\zeta_t^{(t,m)}$ to produce the approximation to (\ref{eq:pred}).
 \item[(iii)] For $m\in\mathcal{M}$, set
 \[
 w_t^{(t,m)}=\Pr(M=m)P_t\left(t+1,m,\zeta_t^{(t,m)}\right)\sum_{s=0}^{t-1}\sum_{m'\in\mathcal{M}} w_t^{(t,m')}\left(\frac{G(t+1-s)-G(t-s)}{1-G(t-s)}\right),
 \]
 and $\zeta_{t+1}^{(t,m)}=u_t(t+1,m,\zeta_t^{(t,m)})$.
 \item[(iv)] Normalise weights, $w_{t+1}^{(s,m)}$.
\end{itemize}
\end{description}
\end{algorithm}

{\bf{Example Revisited}}

We now give details of step (ii) of the algorithm for the piecewise polynomial regression model. Remember $\zeta_t=(\nu_t,\gamma_t,\mathbf{\mu}_t,\mathbf{D}_t)$. For a new segment with $M_{t+1}=1$ we have $\mathbf{\mu}_t=\mathbf{0}$ and $\mathbf{D}_t=\mathbf{D}_0$. We choose $\nu_t$ and $\gamma_t$ to match moments of the predictive distribution of $\sigma_{t+1}^{-2}$. 

Assume $\nu_t^{(s,m')}$ and $\gamma_t^{(s,m')}$ are the first two components of $\zeta_t^{(s,m')}$. Then we solve
\[
\mbox{E}(\sigma_{t+1}^{-2})=\sum_{s=0}^{t-1}\sum_{m'=1}^2 w_t^{(s,m')}\frac{\nu_t^{(s,m')}}{\gamma_t^{(s,m)}}=
\frac{\nu_{t}^{(t,1)}}{\gamma_t^{(t,1)}}.
\]
and
\[
\mbox{E}(\sigma_{t+1}^{-4})=\sum_{s=0}^{t-1}\sum_{m'=1}^2 w_t^{(s,m')}\frac{\nu_t^{(s,m')}(2+\nu_t^{(s,m')})}{(\gamma_t^{(s,m)})^2}=
\frac{\nu_{t}^{(t,1)}(2+\nu_{t}^{(t,1)})}{(\gamma_t^{(t,1)})^2}.
\]
for $\nu_t^{(t,1)}$ and $\gamma_t^{(t,1)}$.

For a new segment with $M_{t+1}=2$, we have identical calculations for $\nu_t^{(t,2)}$ and $\gamma_t^{(t,2)}$. However, in this case we have $\mathbf{\mu}_{t+1}=(\eta,0,0)$ and $\mathbf{D}_{t+1}=\mbox{Diag}(\tau,\delta_1,\delta_2)$ for some $\eta$ and $\tau$ to be calculated. Again we choose values based on matching moments, this time of $\beta_{t+1,0}$.

Let $\Delta_s=(x_{t+1}-x_{s+1})$, and $\mathbf{a}_s=(1,\Delta_s,\Delta_s^2)^T$, then
\[
\mbox{E}(\beta_{t+1,0})=\sum_{s=0}^{t-1}\sum_{m'=1}^2 w_t^{(s,m')} \mathbf{\mu}_t^{s,m'} \mathbf{a}_s =\eta,
\]
and
\[
\mbox{E}(\beta^2_{t+1,0})=\sum_{s=0}^{t-1}\sum_{m'=1}^2 w_t^{(s,m')}\left[ \mathbf{a}_s^T \mathbf{D}^{(s,m'}_t \mathbf{a}_s+ (\mathbf{\mu}_t^{s,m'} \mathbf{a}_s)^2\right] =\eta^2+\tau.
\]

\subsection{Smoothing}

Once we have calculated the filtering distributions for all $t$, we can simulate, backwards in time, the number and position of changepoints, the segment models and parameters, given the full data $y_{1:n}$.

Firstly, we can simulate $(C_n,M_n,\theta_n)$ from (our approximation to) $p(C_n,M_n,\theta_n|y_{1:n})$. These will give us the start of the final segment, together with its model and parameter values. Assume we simulate $C_n=t$, then we will next simulate $(C_t,M_t,\theta_t)$ from
\[ 
p(C_t,M_t,\theta_t|y_{1:n},C_{t+1}=t,C_{t+2:n},M_{t+1:n},\theta_{t+1:n}).
\]
This will give us the start of the penultimate segment, its model and parameter values. We can then repeat this backwards in time until we simulate the first segment for our data.

To perform the simulation we use the fact that
\[
 p(C_t,M_t,\theta_t|y_{1:n},C_{t+1:n},M_{t+1:n},\theta_{t+1:n})=p(C_t,M_t,\theta_t|y_{1:t},C_{t+1},M_{t+1},\theta_{t+1}),
\]
by the conditional independence structure of the model. Thus we have
\begin{eqnarray*}
 \lefteqn{p(C_t=s,M_t=m,\theta_t|y_{1:n},C_{t+1}=t,C_{t+2:n},M_{t+1}=m',M_{t+2:n},\theta_{t+1:n})}\\
&=&p(C_t=s,M_t=m,\theta_t|y_{1:t},C_{t+1}=t,M_{t+1}=m',\theta_{t+1})\\
&\propto& p(C_t=s,M_t=m,\theta_t|y_{1:t})p(C_{t+1}=t,M_{t=1}=m',\theta_{t+1}|C_t=s,M_t=m,\theta_t,y_{1:t}) \\
&\propto&p(C_t=s,M_t=m,\theta_t|y_{1:t})\Pr(C_{t+1}=t|C_t=s)p(\theta_{t+1}|C_{t+1}=t,M_{t+1}=m',C_t=s,M_t=m,\theta_t),
\end{eqnarray*}
where in the final step we have used that the model of a new segment is independent of the model of the preceeding segment.

To simplify notation, let $\mathcal{F}_t=\{y_{t+1:n},C_{t+1:n},M_{t+1:n},\theta_{t+1:n}\}$ denote the future of the process after time $t$. Now substituting $p(C_t=s,M_t=m,\theta_t|y_{1:t})=w_t^{(s,m)}p(\theta_t|\zeta^{(s,m)}_t)$ we get
\begin{eqnarray} 
 \lefteqn{\Pr(C_t=s,M_t=m|y_{1:t},\mathcal{F}_t)\propto w_t^{(s,m)} \Pr(C_{t+1}=t|C_t=s)} \nonumber \\ 
& \times&\int p(\theta_t|\zeta_t^{(s,m)})p(\theta_{t+1}|C_{t+1}=t,M_{t+1}=m',C_t=s,M_t=m,\theta_t) \mbox{d}\theta_t, \label{eq:sm1}
\end{eqnarray}
and
\begin{equation} \label{eq:sm2}
 p(\theta_t|C_t=s,M_t=m,y_{1:t},\mathcal{F}_t) \propto p(\theta_t|\zeta_t^{(s,m)})p(\theta_{t+1}|C_{t+1}=t,M_{t+1}=m',C_t=s,M_t=m,\theta_t).
\end{equation}
We need to be able calculate (or approximate) the integral in (\ref{eq:sm1}) and simulate from (\ref{eq:sm2}) to perform the smoothing.
The full smoothing algorithm is given by Algorithm \ref{alg:smooth}.
\begin{algorithm}
 \caption{Smoothing Algorithm}
\label{alg:smooth}
\begin{description}
 \item[Initiate] \begin{enumerate}
                  \item 
Simulate $(C_n,M_n)$ from the discrete distribution that gives probability $w_n^{(s,m)}$ to the value $(s,m)$. Assuming $(C_n,M_n)=(s,m)$, then simulate $\theta_n$ from $p(\theta_n|\zeta_n^{(s,m)})$.\\
                  \item 
 Set $t=s$,$m'=m$ and $\theta=\theta_n$.
                 \end{enumerate}
\item[While $t>0$]
\begin{enumerate}
 \item For $s=0,\ldots,t-1$ and $m\in\mathcal{M}$ calculate 
\begin{eqnarray*}
\lefteqn{\tilde{w}^{(s,m)}=w_t^{(s,m)}\Pr(C_{t+1}=t|C_t=s)} \\
& \times & \int p(\theta_t|\zeta_t^{(s,m)})p(\theta_{t+1}=\theta|C_{t+1}=t,M_{t+1}=m',C_t=s,M_t=m,\theta_t) \mbox{d}\theta_t
\end{eqnarray*}
\item Simulate $(C_t,M_t)$ from the discrete distribution that gives probability proportional to $\tilde{w}^{(s,m)}$ to the value $(s,m)$. 
\item Assume $(C_t,M_t)=(s,m)$. Simulate $\theta_t$ from the distribution proportional to 
\[
 p(\theta_t|\zeta_t^{(s,m)})p(\theta_{t+1}=\theta|C_{t+1}=t,M_{t+1}=m',C_t=s,M_t=m,\theta_t).
\]
\item Set $t=s$, $m'=m$ and $\theta=\theta_{t+1}$.
\end{enumerate}
\end{description}
\end{algorithm}
 The smoothing algorithm simulates the number and position of the changepoints, and the segment models and parameters. Often more accurate results can be obtained by throwing away the simulated parameter values, and re-simulating these from their conditional distribution given the changepoints and segment models (assuming this distribution is tractable). Such an approach is possible for our piecewise polynomial regression example, and is what we used in the simulation studies later.

We now give details of the calculations involved in the smoothing algorithm for our example.

{\bf Example Revisited}

For our example $\theta_t=(\sigma_t,\beta_t)$. Consider a changepoint at $t$, and $C_t=s$. Define  $\mathbf{h}=(1,\Delta,\Delta^2)$ and $\Delta=(x_{t+1}-x_{s+1})$.  

Firstly consider calculating an integral of the form
\[
 \int p(\theta_t|\zeta)p(\theta_{t+1}|C_{t+1}=t,M_{t+1}=m',C_t=s,M_t=m,\theta_t) \mbox{d}\theta_t,
\]
where $\zeta=(\nu,\gamma,\mu,\mathcal{D})$, for step 1 of Algorithm \ref{alg:smooth}. For $m'=1$ this becomes 
\[
 \mbox{IG}(\sigma_{t+1};\nu/2,\gamma/2)\mbox{N}(\beta_{t+1};\mathbf{0},\sigma_{t+1}^2\mathbf{D}_0),
\]
where $\mbox{IG}(x;a,b)$ denotes the probability density function (pdf) of an inverse-gamma distribution with parameter $a$ and $b$, evaluated at $x$; and $\mbox{N}(\mathbf{x};\eta,\Sigma)$ denotes the pdf of a multivariate normal distribution with mean $\mu$ and variance $\Sigma$, evaluated at $\mathbf{x}$. The first term comes from the fact that $\sigma_{t+1}=\sigma_t$, and second due to the independence of $\beta_{t+1}$ and $\beta_t$. For $m'=2$, the integral becomes
\[
 \mbox{IG}(\sigma_{t+1};\nu/2,\gamma/2)\mbox{N}(\beta_{t+1};\eta,\sigma_{t+1}^2\Sigma),
\]
where $\eta=(\mathbf{h}\mu^T,0,0)$, and $\Sigma=\mbox{Diag}(\mathbf{h}^T\mathbf{D}\mathbf{h},\delta_1,\delta_2)$. Here the conditional density for $\beta_{t+1}$ has changed as now $\beta_{t+1,0}=\mathbf{h}\beta_t$ due to continuity.

Now consider simulating $\theta_t$ from a density proportional to 
\[
  p(\theta_t|\zeta)p(\theta_{t+1}|C_{t+1}=t,M_{t+1}=m',C_t=s,M_t=m,\theta_t),
\]
in step 3 of Algorithm \ref{alg:smooth}. For $m'=1$ we set $\sigma_t=\sigma_{t+1}$, and simulate $\beta_t$ from a multivariate normal distribution with mean $\mu$ and variance $\sigma_{t+1}^2\mathbf{D}$. For $m'=2$ we again set $\sigma_t=\sigma_{t+1}$, but now simulate $\beta_t$ from a multivariate normal distribution with mean $\mu$ and variance $\sigma_{t+1}^2\mathbf{D}$ conditional on $\mathbf{h}\beta_t^T=\beta_{t+1,0}$. Standard results \cite[see e.g.][]{Rue/Held:2005}, gives that we simulate $\beta_t$ from a multivariate normal with mean
\[
\mu-\mathbf{D}\mathbf{h}^T(\mathbf{h}\mathbf{D}\mathbf{h}^T)^{-1}(\mathbf{h}\mu^T-\beta_{t+1,0}) ,
\]
and variance
\[
\sigma_{t+1}^2\left( \mathbf{D}-\mathbf{D}\mathbf{h}^T(\mathbf{h}\mathbf{D}\mathbf{h}^T)^{-1}\mathbf{h}^T\mathbf{D}\right).
\]

\subsection{Resampling} \label{S:Res}

Simulating from the posterior distribution of the number and position of changepoints, and the segment models and parameters, using the filtering and smoothing algorithms has a complexity which is quadratic in $n$. This is due to the number of support points of $(C_t,M_t)$ increasing linearly with $t$.

At the expense of further approximation, we can develop an algorithm whose total computational cost is linear in $n$ via using particle-filter resampling algorithms \cite[e.g.][]{Liu/Chen/Wong:1998,Fearnhead/Clifford:2003} to approximate the distributions of $(C_t,M_t)$ by  discrete distributions with fewer support points. (The resampling procedures ensure that the number of support points in the resulting approximation is bounded by a constant for all $t$.) This was investigated in \cite{Fearnhead/Liu:2007}, who propose two optimal resampling algorithms for changepoint models, and show that substantial computational savings can be obtained with negligible approximation error. 
\section{Simulation Study}

We now evaluate out method through a simulation study using the piecewise quadratic model introduced within our example. We first look at the accuracy of our filtering and smoothing method for simulating from the posterior distribution, and then compare the accuracy of our method to other approaches for curve-fitting. Finally we look at the accuracy of our method at inferring discontinuities in the underlying curve.

In implementing our method we used the filter and smoothing algorithms with the stratified rejection control resampling method of \cite{Fearnhead/Liu:2007}. The threshold parameter within the resampling algorithm was set to $10^{-6}$ \cite[see][for details]{Fearnhead/Liu:2007}. We used the filter and smoothing algorithms to simulate the number and position of changepoints, the value of the observation variance and the model for each segment. Conditioned on these, we then simulated the $\beta$ values associated with each segment from their conditional distribution. 

The filtering and smoothing algorithms were implemented within {\texttt{C++}} and {\texttt{R}}. The computational cost of the algorithms is roughly linear in the number of observations, and to run them on a data set with 4000 data points took of the order of 10 seconds on a desktop PC.

\subsection{Accuracy of the Simulation Method}\label{S:acc}

To test the accuracy of the filtering and smoothing algorithms at drawing samples from the true posterior distribution, we ran a simulation study where we simulated data under the exact model that we used for analysis. We then calculated the posterior quantiles of the true value for $\sigma$ and the value of the underlying curve at each time point. The rationale is that if we could draw from the true posterior, then these posterior quantiles should be uniformly distributed on $[0,1]$. Any inaccuracies in our simulation method will be demonstrated through deviations of the posterior quantiles from such a uniform distribution.

We simulated data for the piecewise-quadratic model with $\mathbf{D}_0=\mbox{Diag}(1,10^2,40^2)$, $p=4/n$ and $\sigma^2=1$. We analysed the data under the model with the same value for $\mathbf{D}_0$ and $p$, but with an improper prior for $\sigma^2$ (equivalent to $\nu=\gamma=0$). To detect any affect that the amount of data had on the performance of our method we simulated 100 data sets for each of $n=256$, $512$ and $1024$. In each case we used equally spaced $x_t$ points in $[0,1]$.

\begin{figure}
 \begin{center}
 \includegraphics[angle=270,scale=0.5]{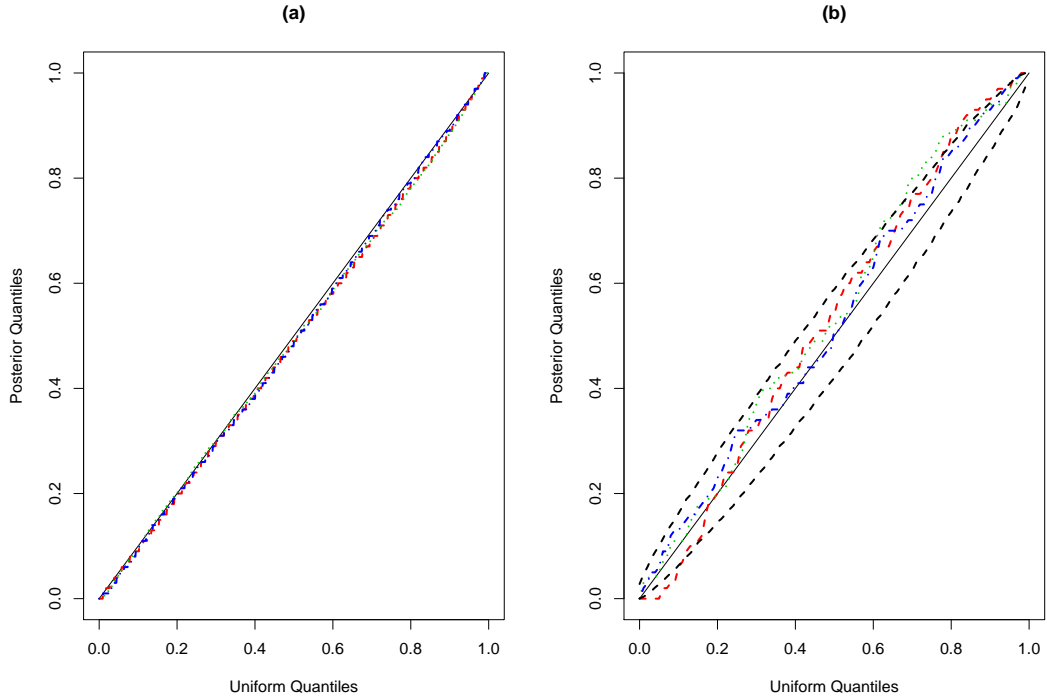}
\caption{\label{Fig:1} Posterior quantile plots of (a) the underlying curve and (b)  $\sigma^2$ for different values of $n$: 256 (red, dashed line), 512 (green dotted line) and 1024 (blue dot-dashed line).  For (b) we give $90\%$ confidence intervals obtained through simulation (black dashed line).}
\end{center}
\end{figure}

Plots of the posterior quantiles are shown in Figure \ref{Fig:1}. In both cases they are close to that expected if they were drawn from the true posterior distribution. The extra smoothness in the plot of posterior quantiles of the underlying curve is due to the larger number of quantiles obtained in this case, $100n$ as we obtain one quantile for each data point. For the posterior quantiles of $\sigma$ we are able to construct confidence intervals, as the posterior quantiles are independent. We notice that the observed quantiles generally lie within the plotted $90\%$ confidence interval. Taken together, these results suggest that negligible error is being introduced by the approximations in our method for simulating from the posterior distribution.

\subsection{Comparison for curve-fitting}

We now look at the accuracy of our piecewise quadratic regression model, together with the new simulation method, for curve-fitting.
Firstly, in order to implement our method we need to choose the prior parameter values. As above we will use the default uninformative prior for $\sigma$. We will assume no prior knowledge of $\mathbf{D}_0$ and $p$, and use an empirical Bayes approach to estimate these hyper-parameters  \cite[as suggested in][]{Fearnhead:2005IEEE}, whereby we estimate their values from the data. We did a preliminary analysis of the data (using default choices for $\mathbf{D}_0$ and $p$), and then estimated $\mathbf{D}_0$ and $p$ from the posterior distribution of the $\beta$s and the number of changepoints. If necessary this could be repeated, with simulation from the posterior given the latest estimates for $\mathbf{D}_0$ and $p$, and new estimates of $\mathbf{D}_0$ and $p$ obtained. 

For our simulation study we chose default value of $p=1/n$ and $\mathbf{D}_0=\mbox{diag}(10,100\times 10^2,1000\times40^2)$. These
are substantially different from the true values used in the simulation (see above). For simplicity we did not repeat the iterative procedure just described. The effect of these choices are discussed below.

We first quantify the accuracy of our method for analysing the same simulated data sets that were used in Section \ref{S:acc}. For a given data set let $z_t$ denote the value of the underlying curve at time $t$ (so observations are $y_t=z_t+\sigma\epsilon_t$ where $\epsilon_t$ is a standard normal random variable). Denote by $\hat{z}_t$ an estimate of $z_t$, then we estimate the accuracy of an estimate of the curve $z_{1:n}$ by the average mean square error
\[
\mbox{MSE}= {\frac{1}{n}\sum_{t=1}^n (z_t-\hat{z}_t)^{2}}.
\]
For our method we use the posterior mean as our estimate of $z_t$. We also look at the mean point-wise coverage of $90\%$ credible (or confidence) intervals for $z_t$. 

For comparison we estimate the underlying curve using wavelets. We implement two wavelet methods, that of \cite{Abramovich/Sapatinas/Silverman:1998} implemented using the {\texttt{BAYES.THR}} function in {\texttt{R}}, and one using complex wavelets \cite[]{Barber/Nason:2004} implemented using the {\texttt{cthresh}} function in {\texttt{R}}. We also constructed wavelet-based confidence intervals \cite[]{Barber/Nason/Silverman:2002} using the {\texttt{wave.band}} function in {\texttt{R}}. (See {\texttt{http://www.stats.bris.ac.uk/$\sim$wavethresh/}} for details of these functions; we used default settings for the {\texttt{R}} functions in all cases.)

\begin{table}
\begin{center}
\begin{tabular}{c|ccc|cc}
    & \multicolumn{3}{c|}{MSE} & \multicolumn{2}{c}{Coverage}  \\
$n$ & New & {\texttt{BAYES.THR}} & {\texttt{cthresh}} & New & {\texttt{wave.band}} \\ \hline
256 & 0.056 & 0.215 &  0.15  & 0.87 & 0.79 \\
512 & 0.027 & 0.138 &  0.093 & 0.87 & 0.79 \\
1024& 0.014 & 0.087 &  0.056 & 0.89 & 0.79 \\
\end{tabular}
\caption{\label{Tab:R1} Mean square error (MSE) and coverage of putative 90\% confidence/credible intervals for our new method, and wavelet based methods.}
\end{center}
\end{table}

Results for the simulated data described in Section \ref{S:acc} are given in Table \ref{Tab:R1}. We notice that the MSE for estimates of the underlying curve is substantially smaller for  our new approach than for either wavelet method. Of the two wavelet methods, the one using complex wavelets gives superior performance. The MSE of our new method halves each time $n$ is doubled, whereas the MSE of the wavelet methods decreases by a smaller proportion each time. Finally, the coverage of our 90\% credible intervals are close to 90\% in each case. The fact that the coverage of the intervals is less than their putative size is likely to be down to errors in estimating the hyperparameters. 

The choice of default starting values for $p$ and $\mathbf{D}_0$, used in the iterative empirical Bayes procedure, has a small effect on the results. For example for $n=256$, repeating the analysis with default values of $p=10/n$ and $p=20/n$ increased mean square error by $0.001$ and $0.004$ respectively. Increasing the default value for $\mathbf{D}_0$, through scaling by a factor of 10 or 100, increased mean square error by $0.007$ and $0.011$ respectively. In these latter cases, our default value is substantially different from the truth, and we do see a non-negligible increase in mean square error. However we can avoid this by repeating the iterative procedure: for example in the last case repeating the procedure just 3 times leads to the same mean square error as reported in Table \ref{Tab:R1}.

The advantage of our method over a wavelet approach for these data is not suprising as the data was simulated under the model assumed by our method. To test robustness of this method to data being simulated from an alternative model, we repeated our simulation study but with data simulated under a piecewise cubic model. For this model we set $\mathbf{D}_0=\mbox{diag}(1,10^2,40^2,d^2)$, and considered the effect that $d$ had. Note that the expected value of the modulus of the cubic co-efficient is $d(2/\pi)^{1/2}$. For simplicity we fixed $n=256$ for all simulations that we carried out.

Results are given in Table \ref{Tab:R2}, again based on 100 simulated data sets for each set of parameters. As expected, as $d$ increases, which corresponds to an increasingly non-quadratic components of the underlying curve, the performance of the new method deteriorates. This is both in terms of the coverage properties of the credible intervals, and the mean square error of estimates of the underlying curve. However for all values of $d$ we considered, the new method still substantially out-performs both wavelet methods in terms of estimating the underlying curve. 

\begin{table}
\begin{center}
\begin{tabular}{c|ccc|cc}
    & \multicolumn{3}{c|}{MSE} & \multicolumn{2}{c}{Coverage}  \\
$d$ & New & {\texttt{BAYES.THR}} & {\texttt{cthresh}} & New & {\texttt{wave.band}} \\ \hline
100 & 0.06 & 0.34 & 0.16 & 0.86 & 0.80 \\
200 & 0.07 & 0.69 & 0.17 & 0.86 & 0.82 \\
400 & 0.11 & 2.45 & 0.18 & 0.84 & 0.86
\end{tabular}
\caption{\label{Tab:R2} Mean square error (MSE) and coverage of putative 90\% confidence/credible intervals for our new method, and wavelet based methods. Data simulated under a piecewise cubic model, with $d$ affecting the size of the cubic co-efficients. All data sets were simulated with $n=100$.}
\end{center}
\end{table}

\begin{figure}
 \begin{center}
 \includegraphics[angle=270,scale=0.65]{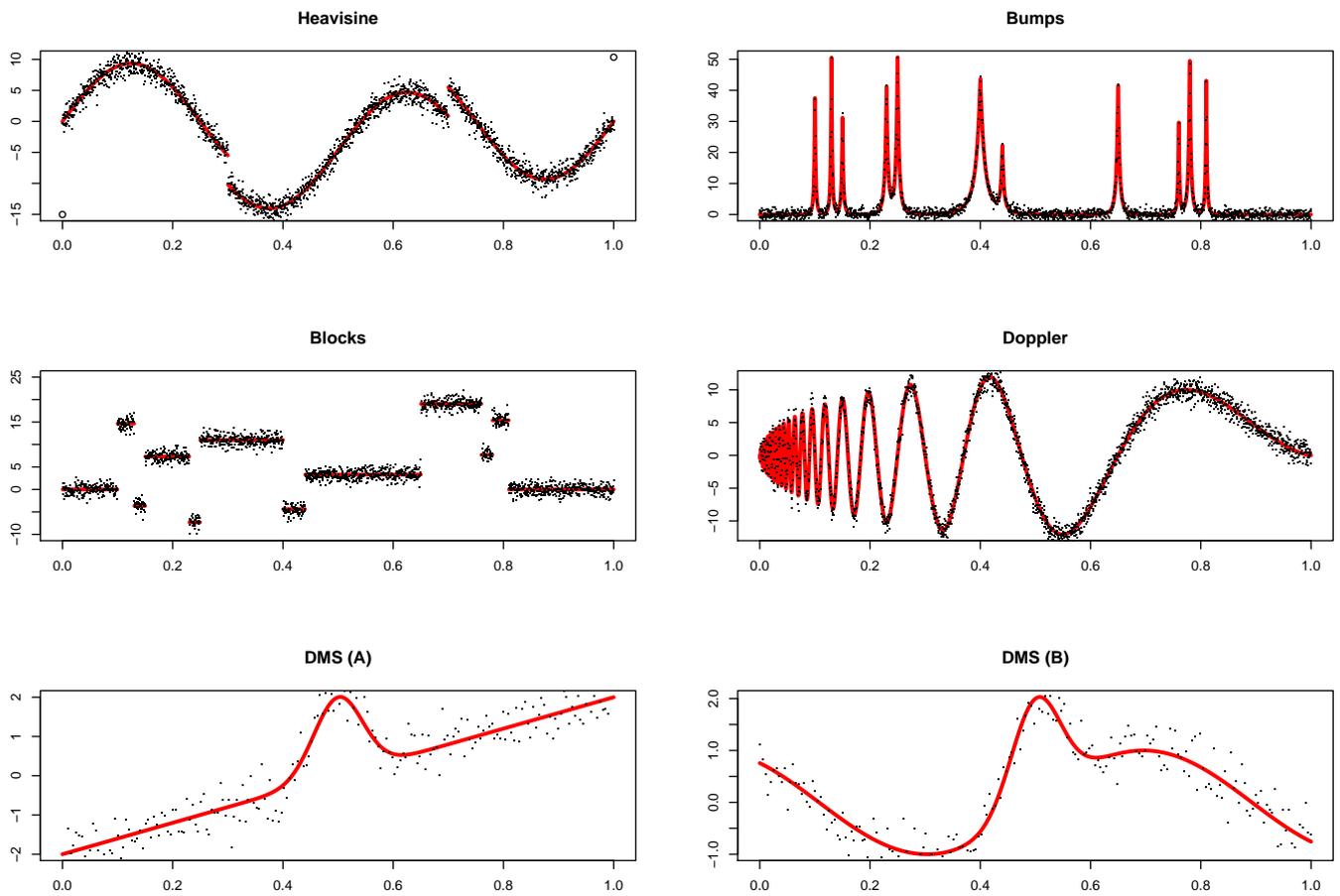}
\caption{\label{Fig:2} Simulated data sets used for comparison with method of \cite{Denison/Mallick/Smith:1998} }
\end{center}
\end{figure}
As a final comparison, we applied our new method to various test data sets from the literature, and compare our method with the published results of \cite{Denison/Mallick/Smith:1998} (henceforth DMS). The test data sets used are shown in Figure \ref{Fig:2}, and consist of the Heavisine, Blocks, Bumps and Doppler signals of \cite{Donoho/Johnstone:1994};  and the smooth function (a) and (b) from \cite{Denison/Mallick/Smith:1998} (denoted DMS A amd DMS B). The method of \cite{Denison/Mallick/Smith:1998} uses a reversible jump MCMC to fit a piecewise cubic function, under continuity and differentiability constraints. The MCMC algorithm samples from an approximation to the posterior, based on approximating the marginal likelihood for each segment. The MCMC procedure takes up to about an order of magnitude longer to analyse the data than our approach.

\begin{table}
\begin{center}
\begin{tabular}{c|ccc|ccc}
    & $n$ & $\sigma$ & SNR & DMS  & NEW  & {\texttt{cthresh}} \\ \hline
Heavisine & 2048 & 1.0 & 7 & 0.033  & 0.022 & 0.032\\
Blocks & 2048 & 1.0 & 7 & 0.170 &  0.016 & 0.116\\
Bumps & 2048 & 1.0 & 7 & 0.167 & 0.318 & 0.100\\
Doppler & 2048 & 1.0 & 7 & 0.135& 0.198 & 0.050\\
DMS A & 200 & 0.4 & 3 & 0.010 & 0.010&\\
DMS B & 200 & 0.3 & 3 & 0.009 & 0.006&\\
\end{tabular}
\caption{\label{Tab:R3} MSE results for 6 test data sets (see Figure \ref{Fig:2}). For each data set we give the number of data points, $n$, the observation error, $\sigma$, and the signal-to-noise ratio.  MSE results for DMS are taken from \cite{Denison/Mallick/Smith:1998}.}
\end{center}
\end{table}

We compare methods based on MSE as before. Results are given in Table \ref{Tab:R3}. Our method does considerably better at estimating the curves which contain discontinuities, as our model allows for discontinuities in the underlying curve. While we do similarly or better on DMS A and DMS B, our method is substantially worse for the Bumps and Doppler data sets. This is due to errors in estimating the peaks in the Bumps data set, and the initial part of the curve in the Doppler data set. In both cases these are where the underlying curve changes most rapidly.  One explanation for this is that using only quadratic polynomials, rather than cubic, makes it harder for our model to fit these parts of the curve. 

The results in \cite{Denison/Mallick/Smith:1998} suggest that the DMS method is more accurate than using wavelets. We investigated this by calculating mean square errors for estimates obtained using the complex wavelet method implemented in {\texttt{cthresh}}. Results are given in Table \ref{Tab:R3} for the four data sets where the number of observations were an integer power of 2 (and thus it is straightforward to apply the wavelet approach). We get different results from \cite{Denison/Mallick/Smith:1998}, with the wavelet approach out-performing the other two approaches for Bumps and Doppler, and out-performing DMS for Blocks.

\subsection{Power at detecting discontinuities}

Finally we look at the power of our method for detecting discontinuities in the underlying curve. Note that it is only our method that can potentially distinguish between changepoints at which the underlying curve may be either continuous or discontinuous. We focus on this feature of our method due to the application of the method we consider in Section \ref{S:WL}.

We used as a basis the continuous curve in DMS B (see Figure \ref{Fig:2}). We then introduced a discontinuity into the curve. If we denote the underlying DMS B curve by $f(x)$ for $x\in[0,1]$, then we introduce a changepoint of size $c$ at point $x_c$ to produce the curve:
\[
 f(x;c,x_c)=\left\{\begin{array}{cl} f(x)-c\sigma &\mbox{ for $x<x_c$,} \\
                    f(x) & \mbox{ for $x\geq x_c$,} \end{array} \right.
\]
where $\sigma^2$ is the variance of the observations. We then simulated data centered on this curve, and look at the posterior probability of a discontinuous changepoint at between $[x_c-0.01,x_c+0.01]$. We repeated this for different values of $c$, $x_c$ and sample size $n$.

\begin{figure}
 \begin{center}
 \includegraphics[angle=270,scale=0.5]{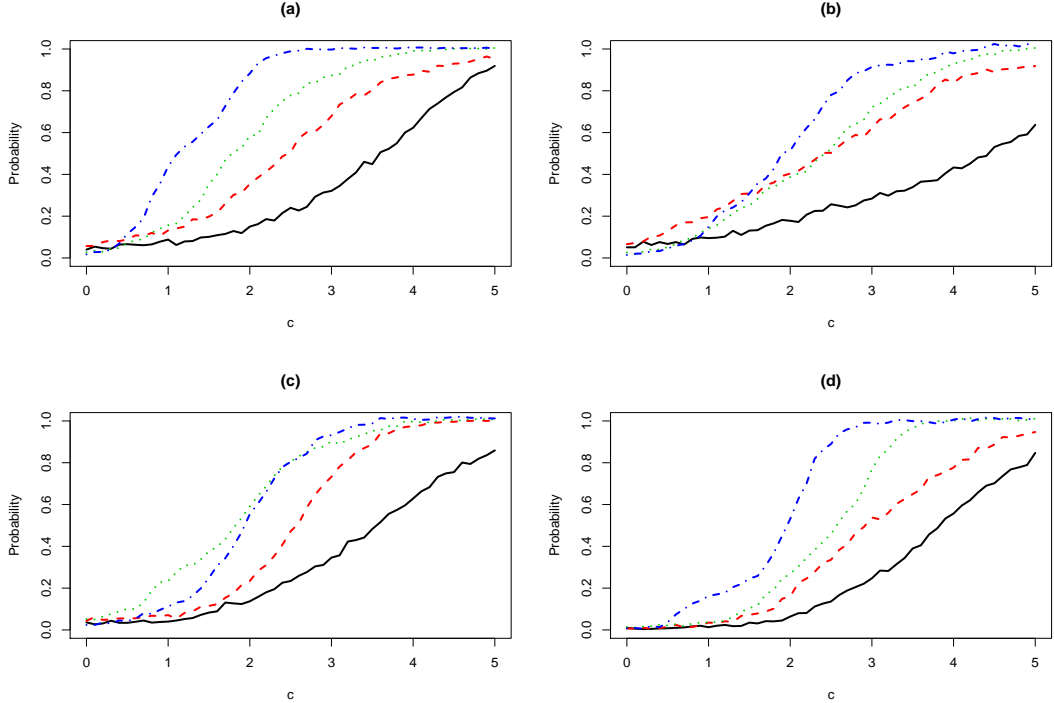}
\caption{\label{Fig:3} The posterior probability of a changepoint within $[x_c-0.01,x_c+0.01]$ for DMS A, for different changepoint positions $x_c$, size of changepoint $c$ and number of observations $n$. Figure (a) is $x_c=0.3$, (b) is $x_c=0.45$, (c) is $x_c=0.6$ and (d) is $x_c=0.7$. For each plot the lines correspond to different values of $n$: $n=100$ (black full line); $n=200$ (red dashed line); $n=400$ (green dotted line); and $n=800$ (blue dot-dashed line).}
\end{center}
\end{figure}

Results are given in Figure \ref{Fig:3}. As expected the posterior probability of a changepoint increases with both $n$ and $c$, and to a lesser extent by the position of the changepoint. The lowest posterior probability of a changepoint occurs when $x_c=0.45$, which is the point at which the gradient of the signal is greatest, and this makes jumps in the signal harder to infer. In general an average posterior probability of a changepoint of greater than 0.5 can occurs with $c>3$ when $n$ is 200 or more; and when $c>2$ and $n=800$.

\section{Well-log Data} \label{S:WL}

\begin{figure}
 \begin{center}
 \includegraphics[angle=270,scale=0.5]{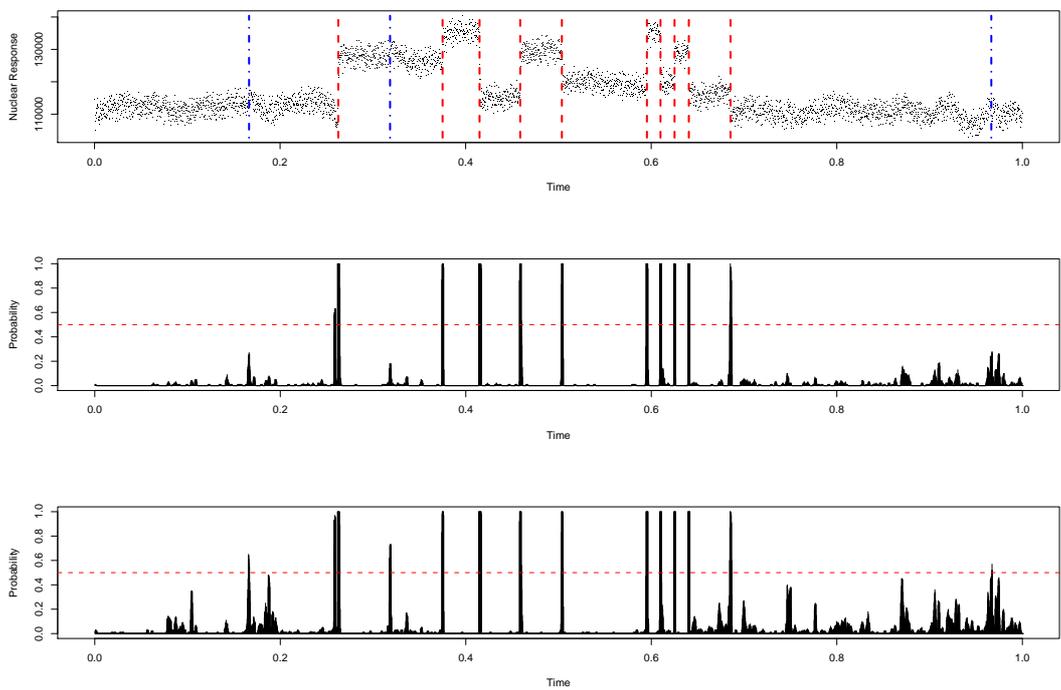}
\caption{\label{Fig:WellLog} (Top) Raw well-log data, with inferred discontinuities (red dashed vertical lines: both models; blue dot-dashed line: model B only). (Middle and Bottom) Posterior probability of discontinuity at time $[t-0.001,t+0.001]$ for model A and B respectively.}
\end{center}
\end{figure}

We now apply our method to analyse the well-log data of \cite{Fitzgerald:1996}. The data is shown in Figure \ref{Fig:WellLog}, and consists of a time-series of measurements of rock as a probe is lowered through a bore-hole in the earth's surface. We have scaled time so that time-series is over the interval $[0,1]$. The underlying signal has a number of abrupt changes, due to the changes in rock strata. It is of interest to locate these abrupt changes in the signal. See \cite{Fitzgerald:1996} and \cite{Fearnhead/Clifford:2003} for further discussion of this data set, and the practical importance of detecting changes in rock strata. Furthermore \cite{Fearnhead/Clifford:2003} discuss the need for online methods for analysing data of this type.

Both \cite{Fitzgerald:1996} and \cite{Fearnhead/Clifford:2003} fit a piecewise constant signal to the data and assume observation error is independent over time. However, \cite{Fearnhead:2006SC} suggests that such a model is inappropriate as it ignores local variation within segments, and fitting such a model results in the detection of too many changepoints. Thus here we will consider analysing the data under our model. The idea is that our model is flexible to allow for variation within rock strata through changepoints at which the underlying signal is continuous. Changes in rock strata will correspond to changepoints at which the underlying signal is discontinuous. Our interest is thus in detecting the position of these discontinuous changepoints.

As in \cite{Fitzgerald:1996} we first remove outliers from the data, and then analyse the data in batch. We consider two analyses, one allowing for the possibility of changepoints at which the underlying signal is either continuous of discontinuous; and the other which only allows changepoints where the underlying signal is discontinuous. The latter mimics the models of \cite{Fitzgerald:1996} and \cite{Fearnhead/Clifford:2003}. We call these models, model A and model B respectively.

Results are given in Figure \ref{Fig:WellLog}. For each model we plot the posterior probability of a discontinuity of the signal in an interval $[t-0.001,t+0.001]$ for different values of $t$. For simplicity we infer a discontinuity whenever this probability is greater than 0.5, and plot the inferred changepoints for the two models. Model B appears to overfit discontinuities in the data (posterior mean number of discontinuities, 30, is nearly twice that for model A), and using our simple procedure for highlighting changepoints, infers an extra three discontinuities in the data -- which by eye look spurious.

\section{Discussion}

We have presented a novel and computationally efficient procedure for Bayesian inference for changepoint models, where there is Markov dependence in the segment parameters. The method is approximate, in that it is based on an approximation to the filtering distribution of parameters associated with a new segment. When used with the resampling idea of Section \ref{S:Res} the resulting algorithm has computational and storage costs that are linear in the number of observations. The simulation results in Section \ref{S:acc}, showed that, for the examples we considered, the error introduced by our approximations were negligible.

One issue with our approach is that it is not simple to quantify the error in the approximation. This is a common issue with approximate methods \cite[see the discussion of][]{Rue/Martino/Chopin:2009}. One approach is to use the approximation we develop as a proposal distribution within an importance sampling method \cite[]{Kim/Shephard/Chib:1998}. This idea is considered in \cite{Liu:2007}, where it is show that the resulting importance sampling approach can be very efficient.

We demonstrated the potential of this new procedure through the fitting of piece-wise quadratic functions. The model we fit allowed for both the possibility of continuity or discontinuity at changepoints. Our simulation studies showed that this model is more accurate at fitting curves that contain discontinuities than the related method of \cite{Denison/Mallick/Smith:1998}. We also showed that it can also perform better at estimating the underlying curve than wavelet procedures, and more accurately characterises the uncertainty in the estimate of the curve. Further advantages of our approach is that it can allow for online inference, and also can allow for inference about the presence and locationdiscontinuities in the underlying signal.

{\bf Appendix}

Here we give details of $P_s(t,m,\zeta)$ and $u_s(t,m,\zeta)$ for the piecewise polynomial regression. 
Now denote $\mathbf{h}=(1,\Delta,\Delta^2)$ where $\Delta=(x_t-x_{s+1})$ (suppressing the dependence on $s$ and $t$).
Then given the most recent changepoint is at time $s$, the mean of the observation at time $t$ is $\mathbf{h}\beta_t^T$.

Remember $\zeta=(\nu,\gamma,\mu,\mathbf{D})$, and define $\zeta'=(\nu',\gamma',\mu',\mathbf{D}')$. Define $e=y_t-\mathbf{h}\mu^T$, 
$Q=\mathbf{h}\mathbf{D}\mathbf{h}^T+1$, and $\mathbf{A}=\mathbf{D}\mathbf{h}^T/Q$.
Then if $\zeta'=u_s(t,m,\zeta)$, we get
\[
 \nu'=\nu+1,
\]
\[
 \gamma'=\gamma+e^2/Q,
\]
\[
 \mu'=\mu+\mathbf{A}e,
\]
\[
 \mathbf{D}'=\mathbf{D}-\mathbf{A}^T\mathbf{A}Q.
\]

Furthermore, let $T_d(x;a,R)$ denote the density of a student's $t$ random variable $d$ degrees of freedonm, and with mean $a$ and scale parameter $R$. Then we have
\[
 P_s(t,m,\zeta)=T_\nu(y_t;\mathbf{h}\mu^T,\gamma Q/\nu).
\]

{\bf Acknowledgements} This work was funded by EPSRC grant GR/T19698. We would like to thank Idris Eckley for helpful discussions.

\bibliographystyle{/home/fearnhea/bib/royal}
\bibliography{/home/fearnhea/bib/thesis,/home/fearnhea/bib/coal}

\end{document}